\documentclass{article}%
\usepackage{amsmath}
\usepackage{hyperref}
\usepackage{amsfonts}
\usepackage{amssymb}
\usepackage{graphicx}
\usepackage{setspace}%
\begin{document}

\title{\ \vspace{-0.8in}\\Do time delay effects explain galactic velocity profiles?}
\author{L. Benkoula, K. Chima, J. Kingsbury, K. Marroquin, M. Yim,\medskip
\and and T. Curtright\thanks{curtright@miami.edu (corresponding author)}\bigskip\\Department of Physics, University of Miami, Coral Gables, FL 33124\medskip\ }
\date{}
\maketitle

\begin{abstract}
Using the gravitoelectromagnetic analogy for weak gravitational fields, we
critique explanations of galactic velocity profiles that invoke time delay
effects (i.e. \textquotedblleft retarded gravity\textquotedblright). \ For
\emph{isotropic,} \emph{time-dependent} matter currents, we show within this
framework that the force exerted on an orbiting body is Newtonian and due only
to the \emph{instantaneous} ambient matter configuration --- \emph{there are}
\emph{no time delay} \emph{effects} in such situations..

\end{abstract}
\tableofcontents

\section{Introduction}

A Keplerian velocity profile exhibits tangential, non-relativistic orbital
speeds $v$ that fall as $1/\sqrt{r}$ for circular orbits of radius $r$ about
an attracting center. \ This profile follows immediately from equating
centripetal and Newtonian gravitational accelerations when an attracting
spherical configuration of total mass $M$ lies entirely within the the orbital
radii of much less massive orbiting bodies. \ A Keplerian profile is observed,
as a very good approximation, for all the planets in
\href{https://en.wikipedia.org/wiki/Kepler%27s_laws_of_planetary_motion#/media/File:Solar_system_orbital_period_vs_semimajor_axis.svg}{our
solar system} \cite{SolarSystem} as well as for those individual planets with
multiple moons and rings, e.g.,
\href{https://en.wikipedia.org/wiki/Moons_of_Saturn}{Saturn} \cite{Saturn}.

However, for more than fifty years observations of galaxies \cite{Vera} and
other large astronomical configurations \cite{Fritz} have shown significant
\href{https://en.wikipedia.org/wiki/Galaxy_rotation_curve}{non-Keplerian
velocity profiles}, often with little or no decrease in $v$ for $r$ larger
than the size of most if not all luminous attracting matter
\cite{VelocityProfileReview}. \ There are two major schools of thought about
why this happens.

One school proposes that a distribution of non-luminous galactic
\textquotedblleft\href{https://en.wikipedia.org/wiki/Dark_matter}{dark
matter}\textquotedblright\ exists and extends far beyond that of luminous
matter \cite{DarkMatterReview}. \ For example, elementary arguments based on
Gauss' law for orbits centered within tenuous but extended spherical
distributions of mass, with only Newtonian gravitation interactions in effect,
show that a satellite on a circular orbit of radius $r$ would have $v\left(
r\right)  =\sqrt{GM\left(  r\right)  /r}$, where $G$ is Newton's constant and
$M\left(  r\right)  $ is the total attracting mass contained within a sphere
of radius $r$. \ If $M\left(  r\right)  $ continues to grow with $r$ then
non-Keplerian profiles are to be expected. \ 

A second school proposes that Newton's laws are
\href{https://en.wikipedia.org/wiki/Modified_Newtonian_dynamics}{drastically
modified}, in particular at large distances where net gravitational attraction
is quite weak, e.g., with accelerations of order $10^{-10}$ meters/second$^{2}%
$ \cite{MONDReview}. \ Of course, as should be expected, there are also some
skeptical scholars who think that both of these schools of thought are wrong,
and that there must be other ways to explain galactic velocity profiles.

An interesting alternative explanation by one such contrarian \cite{Asher}
proposes that non-Keplerian velocity profiles are due to time delay effects in
standard, causal, but otherwise unmodified, Newtonian gravitation theory.
\ This alternative is reconsidered in this essay. \ 

The proposed time delay effects are shown here to be absent in circumstances
where the methods of \cite{Asher}\ would suggest they exist. \ A related class
of exactly solvable examples from electrodynamics does \emph{not} exhibit any
such time delay effects. \ A well-known \emph{mathematical analogy} between
electromagnetism and weak-field gravitation is used to argue the same time
delay effects are also absent in the corresponding gravitational systems.
\ The proposed alternative explanation of velocity profiles fails for these
particular examples because it does not properly account for the physics of
time-dependent mass currents.

\section{Conjectured Time Delay Effects}

The key assumption in \cite{Asher}\ is that the gravitational potential is
causal (i.e. \textquotedblleft retarded\textquotedblright\ in time) and has
the form%
\begin{equation}
U\left(  \overrightarrow{r},t\right)  =-G\int\frac{\rho\left(
\overrightarrow{s},t-\frac{1}{c}\left\vert \overrightarrow{r}%
-\overrightarrow{s}\right\vert \right)  }{\left\vert \overrightarrow{r}%
-\overrightarrow{s}\right\vert }~d^{3}s
\end{equation}
where $\rho$ is the \textquotedblleft time-delayed\textquotedblright\ mass
density \cite{Terminology}. \ This is expanded in powers of the delay time
$\left\vert \overrightarrow{r}-\overrightarrow{s}\right\vert /c$ to obtain an
instantaneous term, an innocuous $\dot{\rho}$ \textquotedblleft zero
force\textquotedblright\ term independent of $\overrightarrow{r}$, and finally
a third $\ddot{\rho}$ term which does not necessarily diminish with distance
even in situations where the time-varying mass density is confined to a
compact region of space.%
\[
U\left(  \overrightarrow{r},t\right)  =-G\int\frac{\rho\left(
\overrightarrow{s},t\right)  }{\left\vert \overrightarrow{r}%
-\overrightarrow{s}\right\vert }~d^{3}s+\frac{G}{c}\int\frac{\partial
}{\partial t}\rho\left(  \overrightarrow{s},t\right)  ~d^{3}s-\frac{G}{2c^{2}%
}\int\frac{\partial^{2}}{\partial t^{2}}\rho\left(  \overrightarrow{s}%
,t\right)  \left\vert \overrightarrow{r}-\overrightarrow{s}\right\vert
~d^{3}s+\cdots
\]
Higher order terms ($\cdots$) in the delay time are neglected based on various
additional hypotheses, a few supporting calculations, and perhaps a bit of
wishful thinking.\ \ Orbits for a non-relativistic body of mass $m$ are then
computed based on the usual force law $\overrightarrow{F}%
=-m\overrightarrow{\nabla}U\left(  \overrightarrow{r},t\right)  $. \ Within
this approximation scheme, because of the $\ddot{\rho}$ term, it is argued in
\cite{Asher} that observed galactic velocity profiles are easily accommodated,
in agreement with experimental data, but without invoking large extended
distributions containing substantial amounts of dark matter.

The key assumption in \cite{Asher} is argued to be valid within the standard
spacetime metric ($g_{\mu\nu}$) framework of
\href{https://en.wikipedia.org/wiki/General_relativity}{general relativity}
\cite{Albert} by considering weak perturbations ($h_{\mu\nu\text{ }}$) of the
Lorentz metric ($\eta_{\mu\nu}$) in the form $g_{\mu\nu}=\eta_{\mu\nu}%
+h_{\mu\nu}$. \ Trace-reversed redefinitions of these perturbations are
governed by well-known linearized field equations%
\begin{equation}
\left(  \frac{1}{c^{2}}\frac{\partial^{2}}{\partial t^{2}}-\nabla^{2}\right)
\overline{h}_{\mu\nu}=-\frac{16\pi G}{c^{4}}~T_{\mu\nu}%
\end{equation}
where $\overline{h}_{\mu\nu}=h_{\mu\nu}-\frac{1}{2}\eta_{\mu\nu}h_{\lambda
}^{\lambda}$ and $T_{\mu\nu}$ is the ambient matter energy-momentum-stress
tensor, and where a particular gauge condition has been chosen, namely,
$\partial^{\mu}\overline{h}_{\mu\nu}=0$. \ In this framework, $U\propto
\overline{h}_{00}$. \ Causal solutions of these linearized field equations are
standard, and correct, as exhibited in the above $U\left(  \overrightarrow{r}%
,t\right)  $. \ 

Unfortunately, the analysis in \cite{Asher}\ contains \emph{some logical
missteps}. \ In particular, the analysis therein discards non-vanishing
contributions due to \emph{mass currents} (contributions analogous to those
due to charge currents in electrodynamics) which should be included in
\emph{both} $\overline{h}_{00}$ and $\overline{h}_{0j}$, and which are often
comparable to $U$ as a consequence of mass conservation for weak gravitational
fields, as shown in the following sections of this essay. \ By discarding
those contributions, the work is in good company \cite{Laplace} but flawed
nevertheless. \ Moreover, a cautious reader would also be wise to worry about
possible \emph{gauge dependence}\ in the analysis of \cite{Asher}. \ In fact,
there is a choice of gauge in general relativity (analogous to the Coulomb
gauge in electromagnetic theory) such that $\overline{h}_{00}$ is
instantaneous without any time delay effects \cite{MTW}. \ These analogies are
more than mere coincidence.

\section{The GEM analogy}

A simple intuitive formalism, well-suited to correct the analysis in
\cite{Asher} and include the effects of mass currents, is provided by
\textquotedblleft GEM\textquotedblright\ ---
\href{https://en.wikipedia.org/wiki/Gravitoelectromagnetism}{Heaviside's
\textquotedblleft gravitoelectromagnetism\textquotedblright} \cite{Oliver}.
\ For weak, time-dependent gravitational fields, with $\left\vert h_{\mu\nu
}\right\vert \ll1$, the leading approximation to general relativity simplifies
to a truncated system of linear equations. \ Here are the basic equations for
that weak-field system:%
\begin{equation}
\overrightarrow{\nabla}\cdot\overrightarrow{\mathbb{E}}=-4\pi G~\rho
\ ,\ \ \ \overrightarrow{\nabla}\times\overrightarrow{\mathbb{E}}%
+\frac{\partial}{\partial t}\overrightarrow{\mathbb{B}}%
=0\ ,\ \ \ \overrightarrow{\nabla}\cdot\overrightarrow{\mathbb{B}%
}=0\ ,\ \ \ \overrightarrow{\nabla}\times\overrightarrow{\mathbb{B}}-\frac
{1}{c^{2}}\frac{\partial}{\partial t}\overrightarrow{\mathbb{E}}=-\frac{4\pi
G}{c^{2}}~\overrightarrow{\mathbb{J}} \label{GEM}%
\end{equation}
where $\overrightarrow{\mathbb{E}}\left(  \overrightarrow{r},t\right)  $ is
the \textquotedblleft gravitoelectric\textquotedblright\ field (in units of
acceleration, $%
\operatorname{m}%
/%
\operatorname{s}%
^{2}$), $\overrightarrow{\mathbb{B}}\left(  \overrightarrow{r},t\right)  $ is
the \textquotedblleft gravitomagnetic\textquotedblright\ field (in inverted
time units, $1/%
\operatorname{s}%
$), $\rho\left(  \overrightarrow{r},t\right)  $ is the mass density (in units
of $%
\operatorname{kg}%
/%
\operatorname{m}%
^{3})$, $\overrightarrow{\mathbb{J}}\left(  \overrightarrow{r},t\right)  $ is
the mass current density (in units of $%
\operatorname{kg}%
/%
\operatorname{m}%
^{2}/%
\operatorname{s}%
$), and of course, $G=6.67\times10^{-11}%
\operatorname{m}%
^{3}/%
\operatorname{kg}%
/%
\operatorname{s}%
^{2}$. \ Local mass conservation for weak-field gravitational systems is
encoded in these first-order partial differential equations, as%
\begin{equation}
\frac{\partial}{\partial t}\rho+\overrightarrow{\nabla}\cdot
\overrightarrow{\mathbb{J}}=0 \label{Conservation}%
\end{equation}

This conservation law and the GEM equations are very familiar in form. \ They
are just relabeled versions of Maxwell's equations for electric and magnetic
fields, $\overrightarrow{E}$ and $\overrightarrow{B}$ (also as first expressed
in terms of the vector calculus by Heaviside), with electric charge and
current densities replaced by mass analogues, and with the SI electromagnetic
constants replaced by $1/\varepsilon_{0}\rightarrow-4\pi G$ and $\mu
_{0}\rightarrow-4\pi G/c^{2}$. \ 

The $\overrightarrow{\mathbb{E}}$ and $\overrightarrow{\mathbb{B}}$\ fields
may be represented by scalar and vector potentials, analogous to those in
electrodynamics, as follows from the first order GEM partial differential
equations in exactly the same way that representations of $\overrightarrow{E}$
and $\overrightarrow{B}$ in terms of potentials follow from Maxwell's
equations. \ Thus, there exist $\mathbb{V}$ and $\overrightarrow{\mathbb{A}}$
such that%
\begin{equation}
\overrightarrow{\mathbb{E}}=-\overrightarrow{\nabla}\mathbb{V-}\frac{\partial
}{\partial t}\overrightarrow{\mathbb{A}}\ ,\ \ \ \overrightarrow{\mathbb{B}%
}=\overrightarrow{\nabla}\times\overrightarrow{\mathbb{A}} \label{Potentials}%
\end{equation}
Although $\mathbb{V}$ and $\overrightarrow{\mathbb{A}}$ vary under gauge
transformations having the standard form, $\mathbb{V}\rightarrow
\mathbb{V}-\frac{\partial}{\partial t}\Omega$, $\overrightarrow{\mathbb{A}%
}\rightarrow\overrightarrow{\mathbb{A}}+\overrightarrow{\nabla}\Omega$, as is
the case in electrodynamics, $\overrightarrow{\mathbb{E}}$ and
$\overrightarrow{\mathbb{B}}$ are invariant under such transformations.

In addition, given the gravitoelectromagnetic fields, the force on a
point-like particle of mass $m$ moving with velocity $\overrightarrow{v}$\ is
given by another, almost familiar equation,%
\begin{equation}
\overrightarrow{F}=m\left(  \overrightarrow{\mathbb{E}}+4\overrightarrow{v}%
\times\overrightarrow{\mathbb{B}}\right)
\end{equation}
where the factor of $4$ is at variance with the usual Lorentz force law, but
correct for motion in a weak gravitational field with the chosen normalization
for $\overrightarrow{\mathbb{B}}$ in the basic field equations. \ Therefore,
with this GEM analogy in hand, it is straightforward to investigate the orbits
of mass $m$ particles in situations where ambient mass currents are present
along with corresponding changes in local mass densities. \ That is to say,
solutions of Maxwell's electromagnetic equations are readily transcribed to
this gravitoelectromagnetic system. \ \bigskip

\textit{Caveat emptor!} \ The GEM formalism is \emph{not} exact except in very
special cases \cite{Robin}.\ \ When there is time-dependence, the
correspondence is only an approximation, albeit a good one for many weak-field
situations \cite{Kip}. \ However, if taken naively at face value, GEM is
misleading when it comes to radiation, since it suggests there is a component
of gravitational radiation with spin one, exactly like electromagnetism.
\ This is \emph{not} true. \ In general relativity the basic plane waves of
gravitational radiation only have maximal helicities $\pm2$ \cite{MTW}. \ The
helicity $0$ and $\pm1$ plane wave solutions are absent as a consequence of
additional gauge invariance. \ That said, GEM does include the effects of both
$\overline{h}_{00}$ and $\overline{h}_{0k}$ in a gauge invariant combination.
\ Absent any radiation fields, this is better for logical consistency than
would be the case if one were to ignore the effects of $\overline{h}_{0k}%
$\ and include just the gauge dependent effects of $\overline{h}_{00}$\ alone.
\ In a chosen gauge these two effects can be comparable, even completely
cancelling, as a consequence of mass conservation for weak gravitational
fields. This is shown to be the case in the\ radiation-free examples to follow.

\section{Exact Results for Isotropic Currents}

For time-dependent, isotropic currents, where $\overrightarrow{\mathbb{J}%
}\left(  \overrightarrow{r},t\right)  \equiv\mathbb{J}\left(  r,t\right)
~\widehat{r}$, solutions of the GEM field equations are given by%
\begin{equation}
\overrightarrow{\mathbb{B}}\left(  \overrightarrow{r},t\right)
=0\ ,\ \ \ \overrightarrow{\mathbb{E}}\left(  \overrightarrow{r},t\right)
=-G\int\frac{\overrightarrow{r}-\overrightarrow{s}}{\left\vert
\overrightarrow{r}-\overrightarrow{s}\right\vert ^{3}}~\rho\left(
\overrightarrow{s},t\right)  d^{3}s \label{Exact}%
\end{equation}
The gravitomagnetic field vanishes for isotropic currents, regardless of the
time dependence in $\mathbb{J}\left(  r,t\right)  $. \ For such currents the
integrand for $\overrightarrow{\mathbb{E}}$\ involves the \emph{instantaneous}
mass density, with absolutely \emph{no time delay} in $\rho$ when
$\overrightarrow{\mathbb{J}}$ is isotropic. \ lf the integral for
$\overrightarrow{\mathbb{E}}$\ is well-defined and if there are no boundary
contributions at spatial infinity, the results in (\ref{Exact})\ are exact
solutions to (\ref{GEM}). \ 

Analogous exact results are obtained for isotropic charge currents in
classical electromagnetism as governed by Maxwell's equations. \ In that
context, there is no electromagnetic radiation for isotropic currents since
$\overrightarrow{B}\left(  \overrightarrow{r},t\right)  =0$. \ There is also
no gravitational radiation produced by isotropic mass currents in the GEM
framework, or otherwise since $d^{3}Q_{jk}/dt^{3}=0$ where $Q_{jk}$ is the
reduced quadrupole moment \cite{MTW}.

To verify that $\overrightarrow{\mathbb{E}}$ and $\overrightarrow{\mathbb{B}}$
given in (\ref{Exact}) are exact solutions, it is straightforward to check
that \emph{all} of the time-dependent GEM equations are satisfied. \ The only
equation that requires any real consideration is the one involving
$\overrightarrow{\nabla}\times\overrightarrow{\mathbb{B}}$. \ When
$\overrightarrow{\mathbb{B}}=0$\ this simplifies to $\frac{\partial}{\partial
t}\overrightarrow{\mathbb{E}}=4\pi G~\overrightarrow{\mathbb{J}}$. \ From the
explicit form for $\overrightarrow{\mathbb{E}}$,%
\begin{equation}
\frac{\partial}{\partial t}\overrightarrow{\mathbb{E}}=-G\int\frac{\left(
\overrightarrow{r}-\overrightarrow{s}\right)  }{\left\vert \overrightarrow{r}%
-\overrightarrow{s}\right\vert ^{3}}\frac{\partial}{\partial t}\rho\left(
\overrightarrow{s},t\right)  ~d^{3}s \label{EdotIntegral}%
\end{equation}
But mass conservation (\ref{Conservation}) for isotropic
$\overrightarrow{\mathbb{J}}\left(  \overrightarrow{r},t\right)  $ implies an
isotropic divergence, $\frac{\partial}{\partial t}\rho\left(
\overrightarrow{s},t\right)  =-\overrightarrow{\nabla_{s}}\cdot
\overrightarrow{\mathbb{J}}\left(  \overrightarrow{s},t\right)  =-\frac
{1}{s^{2}}\frac{\partial}{\partial s}\left(  s^{2}\mathbb{J}\left(
s,t\right)  \right)  \equiv f\left(  s,t\right)  $. \ Thus $\frac{\partial
}{\partial t}\overrightarrow{\mathbb{E}}=-G\int\frac{\left(
\overrightarrow{r}-\overrightarrow{s}\right)  }{\left\vert \overrightarrow{r}%
-\overrightarrow{s}\right\vert ^{3}}~f\left(  s,t\right)  ~d^{3}s$. \ 

The integral expression (\ref{EdotIntegral}) is thereby reduced to an
application of \href{https://en.wikipedia.org/wiki/Shell_theorem}{Newton's
shell theorem}, with the result%
\begin{equation}
\frac{\partial}{\partial t}\overrightarrow{\mathbb{E}}\left(
\overrightarrow{r},t\right)  =-4\pi G~\frac{\widehat{r}}{r^{2}}\int_{0}%
^{r}f\left(  s,t\right)  ~s^{2}ds
\end{equation}
From spherical symmetry and the inverse-square dependence of the integrand in
(\ref{EdotIntegral}), all $f\left(  s,t\right)  =\frac{\partial}{\partial
t}\rho\left(  \overrightarrow{s},t\right)  $ for $s>r$ do not contribute to
$\frac{\partial}{\partial t}\overrightarrow{\mathbb{E}}\left(
\overrightarrow{r},t\right)  $. \ On the other hand, Gauss' law applied to
this final integral over a spherical volume of radius $r$ gives, for isotropic
currents,%
\begin{equation}
\int_{0}^{r}f\left(  s,t\right)  ~s^{2}ds=\frac{-1}{4\pi}\int_{0}%
^{r}\overrightarrow{\nabla_{s}}\cdot\overrightarrow{\mathbb{J}}\left(
\overrightarrow{s},t\right)  d^{3}s=-r^{2}\mathbb{J}\left(  r,t\right)
\end{equation}
This verifies $\frac{\partial}{\partial t}\overrightarrow{\mathbb{E}}=4\pi
G~\overrightarrow{\mathbb{J}}$, thereby establishing the only non-trivial
time-dependent GEM equation that needed to be checked for the solution
(\ref{Exact}).\bigskip

\textit{N.B.} \ This argument to verify the GEM equations does \emph{not}
require\ $\rho\left(  \overrightarrow{r},t\right)  $ to be completely
isotropic. \ Conservation for isotropic $\overrightarrow{\mathbb{J}}\left(
\overrightarrow{r},t\right)  =\mathbb{J}\left(  r,t\right)  ~\widehat{r}$, as
stated in (\ref{Conservation}), only requires the time-\emph{dependent} part
of $\rho\left(  \overrightarrow{r},t\right)  $ to be a function of just $r$
and $t$, without any angular dependence. \ That is to say, the mass density
itself may contain a non-isotropic, \emph{static} component. \ But even if
that is the case, the result for $\overrightarrow{\mathbb{E}}\left(
\overrightarrow{r},t\right)  $\ in (\ref{Exact}) is still correct as a
consequence of the linearity of the GEM system.

\subsection{Velocity Profiles}

It follows that the gravitoelectromagnetic fields due to any $\rho$ and
isotropic $\overrightarrow{\mathbb{J}}$ therefore exert a force on a
point-like particle of mass $m$ that is instantaneous and Newtonian as given
by
\begin{equation}
\overrightarrow{F}\left(  \overrightarrow{r},t\right)
=m~\overrightarrow{\mathbb{E}}\left(  \overrightarrow{r},t\right)
\label{Newtonian}%
\end{equation}
with $\overrightarrow{\mathbb{E}}$ as shown in (\ref{Exact}), completely
without any time delay effects. This contradicts the conclusions drawn in
\cite{Asher}. \ More importantly, for weak-field GEM systems with such
isotropic mass currents, it also shows non-Keplerian velocity profiles require
extended distributions of mass, leading back to the dark matter school of thought.

\subsection{The Truncated Expansion}

Now reconsider the truncated expansion of the causal potentials, as proposed
in \cite{Asher}. \ Even for isotropic charge densities, it is true the
gauge-dependent $\ddot{\rho}$ contribution to $\overrightarrow{\mathbb{E}%
}=-\overrightarrow{\nabla}\mathbb{V-}\frac{\partial}{\partial t}%
\overrightarrow{\mathbb{A}}$ may not vanish. \ However, in Lorenz gauge for
which $\frac{1}{c^{2}}\frac{\partial}{\partial t}\mathbb{V}%
+\overrightarrow{\nabla}\cdot\overrightarrow{\mathbb{A}}=0$, local
conservation of mass (\ref{Conservation}) implies the $\ddot{\rho
}=-\overrightarrow{\nabla}\cdot\overrightarrow{\mathbb{\dot{J}}}$ contribution
in $\overrightarrow{\nabla}\mathbb{V}$ is comparable to the leading
$\overrightarrow{\mathbb{\dot{J}}}$ contribution in $\frac{\partial}{\partial
t}\overrightarrow{\mathbb{A}}$. \ In fact, upon evaluating the angular
integrations \emph{before} the radial integration, for isotropic currents
these two contributions to the gauge-invariant $\overrightarrow{\mathbb{E}}$
\emph{exactly cancel} in Lorenz gauge.%
\begin{equation}
0=\frac{1}{2}\int\ddot{\rho}\left(  s,t\right)  \overrightarrow{\nabla
}\left\vert \overrightarrow{r}-\overrightarrow{s}\right\vert ~d^{3}s+\int%
\frac{\mathbb{\dot{J}}\left(  s,t\right)  \ \widehat{s}}{\left\vert
\overrightarrow{r}-\overrightarrow{s}\right\vert }~d^{3}s
\end{equation}
It is not difficult to show this cancellation persists for isotropic currents
\emph{to all orders} in the time-delayed series expansion of the potentials,
again upon evaluating the angular integrations before the radial integration.
\ That is to say, for all $n\geq0$ in Lorenz gauge,%
\begin{equation}
0=\frac{1}{\left(  n+2\right)  !}\int\frac{\partial^{n+2}}{\partial t^{n+2}%
}\rho\left(  s,t\right)  \overrightarrow{\nabla}\left\vert \overrightarrow{r}%
-\overrightarrow{s}\right\vert ^{n+1}~d^{3}s+\frac{1}{n!}\int\frac
{\partial^{n+1}}{\partial t^{n+1}}\mathbb{J}\left(  s,t\right)  \ \widehat{s}%
\left\vert \overrightarrow{r}-\overrightarrow{s}\right\vert ^{n-1}~d^{3}s
\end{equation}
On the other hand, in Coulomb gauge for which $\overrightarrow{\nabla}%
\cdot\overrightarrow{\mathbb{A}}=0$, the scalar potential $\mathbb{V}$ is
instantaneous just as it is in electrodynamics, while for isotropic currents
$\overrightarrow{\mathbb{A}}$ vanishes. \ All this\ leads once again to
(\ref{Exact}). \ 

Detailed derivations of these features for Lorenz and Coulomb gauge potentials
are given in the Appendix.

\section{Conclusions}

In this essay, selected solvable examples of gravitoelectromagnetism, and
their electrodynamic analogues, were shown \emph{not} to exhibit any of the
time delay (i.e. \textquotedblleft retardation\textquotedblright) effects
suggested by the methods of \cite{Asher}. \ Time delays are absent in the
selected examples when the physical effects of mass currents are properly
taken into account and when gauge invariant fields are computed. \ The
selected examples were chosen to have simplifying features, namely, isotropic
currents subject to weak gravitational fields. \ It is possible that
anisotropy as well as strong fields and relativistic violations of mass
conservation might mitigate these null results. \ For example, near galactic
centers such effects might be significant, but routine order of magnitude
estimates suggest such mitigating effects are probably not important for
orbiting bodies at the outer periphery of galaxies. \ Moreover, the analysis
in \cite{Asher} is not contingent on anisotropy or mass non-conservation.
\ Thus our answer to the question posed in the title is: \emph{No, time delay
effects do not explain galactic velocity profiles}.

For classical electromagnetism, results for $\overrightarrow{E}$ and
$\overrightarrow{B}$ \ analogous to (\ref{Exact}) are exact solutions to exact
equations. \ For GEM, (\ref{Exact}) are exact solutions to approximate
equations. \ However, the main goal of this essay was not to find numerical
results accurate to several decimals, rather it was to expose logical
fallacies that result from keeping only a single gauge-variant term, and then
expanding the causal form of that term in a truncated series of powers of the
time delay. \ That truncated, gauge-variant expansion only gives results it
has been forced to give. \ \emph{Those forced results may have no resemblance
to true solutions}, as shown here for the case of isotropic currents.

\section{Appendix: \ Potentials}

The (gravito) electric and magnetic fields are expressed here in terms of
potentials as obtained in two well-known gauges: \ Lorenz and Coulomb.
\ Throughout the analysis $\rho$ and $\overrightarrow{J}$ are assumed to be
well-behaved such that all integrals exist, and are assumed to fall-off
sufficiently rapidly for large distances such that integrations by parts
produce no surface contributions at spatial infinity, for all times. \ Given
these assumptions, the results to follow are \emph{exact} solutions to
Maxwell's equations. \ Suitably transcribed, they are also exact solutions to
the GEM equations, but only approximate descriptions of weak-field
gravitational systems since GEM is not exact. \ For notational simplicity, in
this Appendix we do not distinguish gravitoelectromagnetic quantities from
their electromagnetic counterparts. $\ $The results for $\overrightarrow{E}$
\& $\overrightarrow{B}$ transcribe in an obvious way to those for
$\overrightarrow{\mathbb{E}}$ \& $\overrightarrow{\mathbb{B}}$.

\subsection*{Lorenz gauge}

In this gauge isotropic current flow leads to causal scalar and vector
potentials that are both non-vanishing. \ Nevertheless, for such currents
there is a complete cancellation of time delay effects in $\overrightarrow{E}%
=-\overrightarrow{\nabla}V-\partial\overrightarrow{A}/\partial t$. \ 

\subsubsection*{The Causal Potentials}

The time-delayed scalar and vector potentials in Lorenz gauges are%
\[
V\left(  \overrightarrow{r},t\right)  =\frac{1}{4\pi\varepsilon_{0}}\int%
\frac{\rho\left(  \overrightarrow{s},t-\frac{1}{c}\left\vert
\overrightarrow{r}-\overrightarrow{s}\right\vert \right)  }{\left\vert
\overrightarrow{r}-\overrightarrow{s}\right\vert }~d^{3}%
s\ ,\ \ \ \overrightarrow{A}\left(  \overrightarrow{r},t\right)  =\frac
{\mu_{0}}{4\pi}\int\frac{\overrightarrow{J}\left(  \overrightarrow{s}%
,t-\frac{1}{c}\left\vert \overrightarrow{r}-\overrightarrow{s}\right\vert
\right)  }{\left\vert \overrightarrow{r}-\overrightarrow{s}\right\vert }%
~d^{3}s
\]
leading to the gauge invariant (gravito) electric and magnetic fields%
\[
\overrightarrow{E}\left(  \overrightarrow{r},t\right)
=-\overrightarrow{\nabla}V\left(  \overrightarrow{r},t\right)  -\frac
{\partial}{\partial t}\overrightarrow{A}\left(  \overrightarrow{r},t\right)
\ ,\ \ \ \overrightarrow{B}\left(  \overrightarrow{r},t\right)
=\overrightarrow{\nabla}\times\overrightarrow{A}\left(  \overrightarrow{r}%
,t\right)
\]

\subsubsection*{The Causal Series}

Expanding the potentials as a formal series in time derivatives\footnote{To
ensure that each term in the series is well-defined, the $n^{th}$ time
derivatives of $\rho$ and $J$ must vanish faster than $1/s^{n+3}$ as
$s\rightarrow\infty$ so that the radial integrations are finite. \ For the
purposes of this Appendix, it is assumed that these conditions hold for all
$n$. \ For example, the charge and current densities could fall exponentially
to zero, or else they could vanish exactly outside some large radius.}
\[
V\left(  \overrightarrow{r},t\right)  =\frac{1}{4\pi\varepsilon_{0}}\sum
_{n=0}^{\infty}\frac{1}{n!}\left(  \frac{-1}{c}\right)  ^{n}\int\left\vert
\overrightarrow{r}-\overrightarrow{s}\right\vert ^{n-1}\frac{\partial^{n}%
}{\partial t^{n}}\rho\left(  \overrightarrow{s},t\right)  ~d^{3}s
\]%
\[
\overrightarrow{A}\left(  \overrightarrow{r},t\right)  =\frac{\mu_{0}}{4\pi
}\sum_{n=0}^{\infty}\frac{1}{n!}\left(  \frac{-1}{c}\right)  ^{n}%
\int\left\vert \overrightarrow{r}-\overrightarrow{s}\right\vert ^{n-1}%
\frac{\partial^{n}}{\partial t^{n}}\overrightarrow{J}\left(
\overrightarrow{s},t\right)  ~d^{3}s
\]
yields a corresponding expression for the electric field, with an
instantaneous \textquotedblleft Coulombic\textquotedblright\ leading term,
\begin{align*}
\overrightarrow{E}\left(  \overrightarrow{r},t\right)   &  =\frac{1}%
{4\pi\varepsilon_{0}}\int\frac{\left(  \overrightarrow{r}-\overrightarrow{s}%
\right)  \rho\left(  \overrightarrow{s},t\right)  }{\left\vert
\overrightarrow{r}-\overrightarrow{s}\right\vert ^{3}}~d^{3}s\ \\
&  -\frac{1}{4\pi\varepsilon_{0}}\ \sum_{n=2}^{\infty}\frac{1}{n!}\left(
\frac{-1}{c}\right)  ^{n}\int\overrightarrow{\nabla_{r}}\left\vert
\overrightarrow{r}-\overrightarrow{s}\right\vert ^{n-1}\frac{\partial^{n}%
}{\partial t^{n}}\rho\left(  \overrightarrow{s},t\right)  ~d^{3}s\\
&  -\frac{\mu_{0}}{4\pi}\sum_{n=0}^{\infty}\frac{1}{n!}\left(  \frac{-1}%
{c}\right)  ^{n}\int\left\vert \overrightarrow{r}-\overrightarrow{s}%
\right\vert ^{n-1}\frac{\partial^{n+1}}{\partial t^{n+1}}\overrightarrow{J}%
\left(  \overrightarrow{s},t\right)  ~d^{3}s
\end{align*}
Note the $n=1$ term in $V$ has no dependence on $\overrightarrow{r}$, hence no
gradient. \ 

\subsubsection*{Charge and/or Mass Conservation}

If $\rho$ is the electric charge (or mass) density and $\overrightarrow{J}$ is
the electric charge (or mass) current density, conservation of charge (or
mass) is expressed locally as%
\[
\frac{\partial}{\partial t}\rho\left(  \overrightarrow{r},t\right)
=-\overrightarrow{\nabla}\cdot\overrightarrow{J}\left(  \overrightarrow{r}%
,t\right)
\]
This leads to a series for the causal electric field where the time
derivatives involve only $\overrightarrow{J}$. \ Shifting the summation index
in the $\rho\left(  \overrightarrow{s},t\right)  $ series, supplanting
$\partial^{n+1}\rho/\partial t^{n+1}$ with $-\partial^{n}%
\overrightarrow{\nabla}\cdot\overrightarrow{J}\left(  \overrightarrow{r}%
,t\right)  /\partial t^{n}$, and combining terms gives
\begin{align*}
\overrightarrow{E}\left(  \overrightarrow{r},t\right)   &  =\frac{1}%
{4\pi\varepsilon_{0}}\int\frac{\overrightarrow{r}-\overrightarrow{s}%
}{\left\vert \overrightarrow{r}-\overrightarrow{s}\right\vert ^{3}}%
~\rho\left(  \overrightarrow{s},t\right)  d^{3}s-\frac{\mu_{0}}{4\pi}%
\sum_{n=0}^{\infty}\frac{1}{n!}\left(  \frac{-1}{c}\right)  ^{n}\frac
{\partial^{n+1}}{\partial t^{n+1}}~\overrightarrow{f_{n}}\left(
\overrightarrow{r},t\right) \\
\overrightarrow{f_{n}}\left(  \overrightarrow{r},t\right)   &  \equiv
\int\left\vert \overrightarrow{r}-\overrightarrow{s}\right\vert ^{n-1}\left(
\overrightarrow{J}\left(  \overrightarrow{s},t\right)  -\frac{1}{n+2}\left(
\overrightarrow{r}-\overrightarrow{s}\right)  \overrightarrow{\nabla_{s}}%
\cdot\overrightarrow{J}\left(  \overrightarrow{s},t\right)  \right)  ~d^{3}s
\end{align*}
With an assumption of null boundary contributions at spatial infinity,
integrating by parts then gives%
\begin{gather*}
\overrightarrow{f_{n}}\left(  \overrightarrow{r},t\right)  =\int\left(
\overrightarrow{J}\left(  \overrightarrow{s},t\right)  \left\vert
\overrightarrow{r}-\overrightarrow{s}\right\vert ^{n-1}+\frac{1}%
{n+2}\overrightarrow{J}\left(  \overrightarrow{s},t\right)  \cdot
\overrightarrow{\nabla_{s}}\left(  \left(  \overrightarrow{r}%
-\overrightarrow{s}\right)  \left\vert \overrightarrow{r}-\overrightarrow{s}%
\right\vert ^{n-1}\right)  \right)  ~d^{3}s\\
=\frac{1}{n+2}\int\left(  \left(  n+1\right)  \overrightarrow{J}\left(
\overrightarrow{s},t\right)  -\left(  n-1\right)  \frac{\left(
\overrightarrow{r}-\overrightarrow{s}\right)  }{\left\vert \overrightarrow{r}%
-\overrightarrow{s}\right\vert ^{2}}\left(  \overrightarrow{r}%
-\overrightarrow{s}\right)  \cdot\overrightarrow{J}\left(  \overrightarrow{s}%
,t\right)  \right)  \left\vert \overrightarrow{r}-\overrightarrow{s}%
\right\vert ^{n-1}~d^{3}s
\end{gather*}

\subsubsection*{Order-by-order Cancellation of Terms}

For isotropic currents, where $\overrightarrow{J}\left(  \overrightarrow{r}%
,t\right)  =J\left(  r,t\right)  ~\widehat{r}$, rotational invariance implies%
\[
\overrightarrow{f_{n}}\left(  \overrightarrow{r},t\right)  =f_{n}\left(
r,t\right)  ~\widehat{r}%
\]
with the above integration then reducing to
\begin{gather*}
f_{n}\left(  r,t\right)  =\frac{1}{n+2}\int\left(  \left(  n+1\right)
\widehat{r}\cdot\widehat{s}-\left(  n-1\right)  \frac{\widehat{r}\cdot\left(
\overrightarrow{r}-\overrightarrow{s}\right)  }{\left\vert \overrightarrow{r}%
-\overrightarrow{s}\right\vert ^{2}}\left(  \overrightarrow{r}%
-\overrightarrow{s}\right)  \cdot\widehat{s}\right)  J\left(  s,t\right)
\left\vert \overrightarrow{r}-\overrightarrow{s}\right\vert ^{n-1}~d^{3}s\\
=\tfrac{2\pi\left(  n+1\right)  }{n+2}\int_{0}^{\infty}ds~s^{2}J\left(
s,t\right)  \int_{0}^{\pi}\left(  \cos\theta-\tfrac{n-1}{n+1}\tfrac{\left(
r-s\cos\theta\right)  \left(  r\cos\theta-s\right)  }{r^{2}+s^{2}%
-2rs\cos\theta}\right)  \left(  \sqrt{r^{2}+s^{2}-2rs\cos\theta}\right)
^{n-1}\sin\theta d\theta
\end{gather*}
Changing variables to $u\equiv\cos\theta$, the remaining angular integration
simplifies to reveal \emph{a vanishing contribution} for \emph{all} $n$.%
\begin{align*}
&  \int_{-1}^{1}\left(  _{\ }\left(  n+1\right)  \left(  r^{2}+s^{2}%
-2rsu\right)  u-\left(  n-1\right)  \left(  r-su\right)  \left(  ru-s\right)
\right)  \left(  \sqrt{r^{2}+s^{2}-2rsu}\right)  ^{n-3}du\\
&  =-\int_{-1}^{1}\frac{d}{du}\left(  \left(  1-u^{2}\right)  \left(
\sqrt{r^{2}+s^{2}-2rsu}\right)  ^{n-1}\right)  du=0
\end{align*}
Therefore, for isotropic current flow, the only surviving contribution in the
causal electric field series is the leading, \emph{instantaneous} term.
\[
\overrightarrow{E}\left(  \overrightarrow{r},t\right)  =\frac{1}%
{4\pi\varepsilon_{0}}\int\frac{\overrightarrow{r}-\overrightarrow{s}%
}{\left\vert \overrightarrow{r}-\overrightarrow{s}\right\vert ^{3}}%
~\rho\left(  \overrightarrow{s},t\right)  d^{3}s
\]
This agrees with the exact result in the main text, Eqn. (\ref{Exact}). \ 

Also note, by rotational invariance, each term in the series for
$\overrightarrow{A}$ is radial and otherwise independent of direction, i.e.
$\int\left\vert \overrightarrow{r}-\overrightarrow{s}\right\vert ^{n-1}%
\frac{\partial^{n}}{\partial t^{n}}\overrightarrow{J}\left(
\overrightarrow{s},t\right)  ~d^{3}s=g_{n}\left(  r,t\right)  ~\widehat{r}$.
\ Thus each term is curl-free, confirming that each term in the series
contributes vanishing $\overrightarrow{B}\left(  \overrightarrow{r},t\right)
$.

\subsection*{Coulomb gauge}

In this gauge isotropic current flow leads to a non-vanishing instantaneous
scalar potential $V$, but to a vanishing causal vector potential
$\overrightarrow{A}$. \ Consequently, for such currents there are again no
time delay effects in $\overrightarrow{E}=-\overrightarrow{\nabla}%
V-\partial\overrightarrow{A}/\partial t$. \ 

\subsubsection*{Instantaneous scalar potential}

In Coulomb gauge, for which $\overrightarrow{\nabla}\cdot\overrightarrow{A}%
\left(  \overrightarrow{r},t\right)  =0$, the scalar potential is always given
by a spatial integral of the charge density without any time delay, namely,%
\[
V\left(  \overrightarrow{r},t\right)  =\frac{1}{4\pi\varepsilon_{0}}\int%
\frac{\rho\left(  \overrightarrow{u},t\right)  }{\left\vert \overrightarrow{r}%
-\overrightarrow{u}\right\vert }~d^{3}u
\]
This contributes an instantaneous component to the electric field, as given by%
\[
-\overrightarrow{\nabla}V\left(  \overrightarrow{r},t\right)  =\frac{1}%
{4\pi\varepsilon_{0}}\int\frac{\overrightarrow{r}-\overrightarrow{u}%
}{\left\vert \overrightarrow{r}-\overrightarrow{u}\right\vert ^{3}}%
~\rho\left(  \overrightarrow{u},t\right)  d^{3}u
\]
The time-derivative of $V$\ is likewise given by%
\[
\tfrac{\partial}{\partial t}V\left(  \overrightarrow{r},t\right)  =\tfrac
{1}{4\pi\varepsilon_{0}}\int\tfrac{\tfrac{\partial}{\partial t}\rho\left(
\overrightarrow{u},t\right)  }{\left\vert \overrightarrow{r}%
-\overrightarrow{u}\right\vert }~d^{3}u=\tfrac{-1}{4\pi\varepsilon_{0}}%
\int\tfrac{\overrightarrow{\nabla_{u}}\cdot\overrightarrow{J}\left(
\overrightarrow{u},t\right)  }{\left\vert \overrightarrow{r}%
-\overrightarrow{u}\right\vert }~d^{3}u=\tfrac{-1}{4\pi\varepsilon_{0}}%
\int\overrightarrow{J}\left(  \overrightarrow{u},t\right)  \cdot
\overrightarrow{\nabla_{r}}\left(  \tfrac{1}{\left\vert \overrightarrow{r}%
-\overrightarrow{u}\right\vert }\right)  ~d^{3}u
\]
upon using local charge conservation and integrating by parts with the
assumption of null boundary contributions at spatial infinity.

\subsubsection*{Vanishing causal vector potential}

The vector potential in Coulomb gauge obeys the wave equation with an
additional source term involving $V$, namely,
\[
\left(  \frac{1}{c^{2}}\frac{\partial^{2}}{\partial t^{2}}-\nabla^{2}\right)
\overrightarrow{A}\left(  \overrightarrow{r},t\right)  =\mu_{0}%
\overrightarrow{J}\left(  \overrightarrow{r},t\right)  -\mu_{0}\varepsilon
_{0}\frac{\partial}{\partial t}\overrightarrow{\nabla}V\left(
\overrightarrow{r},t\right)
\]
The causal solution for $\overrightarrow{A}$\ is%
\begin{align*}
\overrightarrow{A}\left(  \overrightarrow{r},t\right)   &  =\frac{\mu_{0}%
}{4\pi}\int\left(  \overrightarrow{J}\left(  \overrightarrow{s},t_{<}\right)
-\varepsilon_{0}\frac{\partial}{\partial t}\overrightarrow{\nabla_{s}}V\left(
\overrightarrow{s},t_{<}\right)  \right)  \frac{1}{\left\vert
\overrightarrow{r}-\overrightarrow{s}\right\vert }~d^{3}s\\
&  =\frac{\mu_{0}}{4\pi}\int\left(  \overrightarrow{J}\left(
\overrightarrow{s},t_{<}\right)  +\frac{1}{4\pi}~\int\frac{1}{\left\vert
\overrightarrow{s}-\overrightarrow{u}\right\vert }\overrightarrow{\nabla_{u}%
}\left(  \overrightarrow{\nabla_{u}}\cdot\overrightarrow{J}\left(
\overrightarrow{u},t_{<}\right)  \right)  ~d^{3}u\right)  \frac{1}{\left\vert
\overrightarrow{r}-\overrightarrow{s}\right\vert }~d^{3}s
\end{align*}
with $t_{<}=t-\frac{1}{c}\left\vert \overrightarrow{r}-\overrightarrow{s}%
\right\vert $, where local charge conservation was used in the 2nd line to
replace $\frac{\partial}{\partial t}\rho\left(  \overrightarrow{u}%
,t_{<}\right)  =-\overrightarrow{\nabla_{u}}\cdot\overrightarrow{J}\left(
\overrightarrow{u},t_{<}\right)  $. \ (Note that $t_{<}$ does not depend on
$\overrightarrow{u}$.) \ The contribution to $\overrightarrow{A}$ from
$\int\overrightarrow{J}\left(  \overrightarrow{s},t_{<}\right)  /\left\vert
\overrightarrow{r}-\overrightarrow{s}\right\vert ~d^{3}s$\ can also be written
as a double spatial integral using%
\[
\overrightarrow{J}\left(  \overrightarrow{s},t_{<}\right)  =\int%
\overrightarrow{J}\left(  \overrightarrow{s},t_{<}\right)  \delta^{3}\left(
\overrightarrow{s}-\overrightarrow{u}\right)  d^{3}u=-\int\overrightarrow{J}%
\left(  \overrightarrow{s},t_{<}\right)  \nabla_{u}^{2}\left(  \frac{1}%
{4\pi\left\vert \overrightarrow{s}-\overrightarrow{u}\right\vert }\right)
d^{3}u
\]
and integrating by parts, again with null boundary conditions at spatial
infinity. \ But, in so doing, it is helpful to retain the time delay in the
form $t_{<}=t-\frac{1}{c}\left\vert \overrightarrow{r}-\overrightarrow{s}%
\right\vert $ so that $t_{<}$\ does not depend on $\overrightarrow{u}$. \ Then
$\overrightarrow{\nabla_{u}}\times\left(  \overrightarrow{\nabla_{u}}%
\times\overrightarrow{J}\left(  \overrightarrow{u},t_{<}\right)  \right)
=\overrightarrow{\nabla_{u}}\left(  \overrightarrow{\nabla_{u}}\cdot
\overrightarrow{J}\left(  \overrightarrow{u},t_{<}\right)  \right)
-\nabla_{u}^{2}\overrightarrow{J}\left(  \overrightarrow{u},t_{<}\right)  $
where the spatial derivatives do not act on $t_{<}$. \ This leads to%
\[
\overrightarrow{A}\left(  \overrightarrow{r},t\right)  =\frac{\mu_{0}}%
{16\pi^{2}}\int\int\frac{\overrightarrow{\nabla_{u}}\times\left(
\overrightarrow{\nabla_{u}}\times\overrightarrow{J}\left(  \overrightarrow{u}%
,t_{<}\right)  \right)  }{\left\vert \overrightarrow{r}-\overrightarrow{s}%
\right\vert \left\vert \overrightarrow{s}-\overrightarrow{u}\right\vert
}~d^{3}u~d^{3}s\text{ \ \ }%
\]
with $t_{<}=t-\frac{1}{c}\left\vert \overrightarrow{r}-\overrightarrow{s}%
\right\vert $ as above. \ 

For situations with isotropic current flow the current is curl-free,
$\overrightarrow{\nabla_{u}}\times\overrightarrow{J}\left(  \overrightarrow{u}%
,t_{<}\right)  =\overrightarrow{\nabla_{u}}\times\left(  \widehat{u}J\left(
u,t-\frac{1}{c}\left\vert \overrightarrow{r}-\overrightarrow{s}\right\vert
\right)  \right)  =0$. \ Therefore, in Coulomb gauge for localized isotropic
source currents,
\[
\overrightarrow{A}\left(  \overrightarrow{r},t\right)  =0
\]
and the electric field is given entirely by the instantaneous gradient of $V$,
as in (\ref{Exact}), once again without any time delay effects.

\end{document}